\documentclass[aps,pra,reprint,twocolumn,showpacs,showkeys,floatfix,nobibnotes]{revtex4-1}

\usepackage{hyperref}
\hypersetup{colorlinks=true, urlcolor=blue, citecolor=blue, linkcolor=blue}
\usepackage{graphicx}
\usepackage{subfigure}
\usepackage{epstopdf}
\usepackage{lastpage}
\usepackage{braket}

\begin{document}
\title{Two-mode photon-added entangled coherent states and their entanglement properties}

\author{Arpita Chatterjee}
\email{Email id: arpita.sps@gmail.com}
\affiliation{Department of Mathematics, J. C. Bose University of Science and Technology, YMCA, Faridabad 121006, Haryana, India}

\date{\today}

\begin{abstract}
An entangled quantum state is considered by applying a local photon excitation to each mode of an entangled coherent state. The entanglement property is investigated in terms of the entropy of entanglement. It is shown that applying a photon addition can improve the amount of entanglement. It is also examined that in a specific region of parameters, the state $\ket{\psi_1^-(\alpha, m, n)}$ is least entangled when photon excitation is minimum. We study the statistical properties of such states by employing the quasi-probability functions.
\end{abstract}

\pacs{42.50.-p, 42.50.Ct, 42.50.Pq}
%\pacs{42.50.Dv, 42.50.Ct}
%42.50.Ct
%(42.50.-p=Quantum optics, 42.50.Ct=Quantum description of interaction of light and matter; related experiments,
%42.50.Dv=Quantum state engineering and measurements)

%\noindent{\it Keywords}: lower- and higher-order nonclassicality, superposed state

%\submitto{\JPB}
%\maketitle
\keywords{photon-added entangled coherent state, entanglement, $P$ function, $Q$ function}

%\pacs{42.50.-p, 42.50.Ct, 42.50.Pq}
%(42.50.-p=Quantum optics, 42.50.Ct=Quantum description of interaction of light and matter; related experiments,
%42.50.Pq=Cavity quantum electrodynamics; micromasers)
\maketitle

%\footnotetext{Corresponding author, Tel.: +91-11-8527635535,\\Electronic
%address: mailtoarpita@rediffmail.com}

\section{Introduction}

Quantum information processing uses the inherent properties of quantum systems like the entanglement which has been widely considered as a useful resource to perform quantum operations, universal quantum computing and quantum communications \cite{serna}. Most of the entangled states, which are important from the quantum computing perspective, violate Bell-type inequalities \cite{bell}, that means the existence of such states cannot be explained by any local hidden-variable theory. The concept of entanglement was introduced by Einstein et al. \cite{ein} who designed a two-particle state that was strongly entangled both in position and momentum space. For instance, the two-mode squeezed vacuum state exhibits
quantum entanglement between the idle mode and the signal mode, and is often applied as an entangled resource \cite{jiani} for quantum dense coding \cite{ban}.

An entangled coherent state (ECS) \cite{barry1, barry2} can be typified as two-mode continuous-variable states, which are very efficient in both generating and manipulating quantum information protocols \cite{vaid}. In addition, continuous-variable entangled states play a crucial role in performing quantum teleportation \cite{van}, quantum computation \cite{jeong1}, entanglement purification \cite{jeong2}, quantum error corrections \cite{glan} etc. Moreover, a number of theoretical schemes have been proposed to produce ECS in cavity fields (\cite{gerry} and references therein). Therefore, the study of entangled coherent states is of much interest, as coherent states are macroscopic and simplest classical-like continuous-variable states that can be easily obtained from available laser sources. Recently, a new entangled quantum state is introduced by applying local coherent superposition (CS) ($r a^\dagger+ta$) of photon addition and subtraction to each mode of an even entangled coherent state (EECS) \cite{jiani}. It is found that single- and two-mode CS operations can improve the EPR correlation of the EECS in a big ($> 0.88$) and small ($< 0.52$) region of amplitude, respectively. Zhou et al. \cite{zhou} proposed two types of two-mode excited entangled coherent states (TMEECSs) $\ket{\psi_{\pm}(\alpha, m, n)}$, and investigated the influence of photon excitations on quantum entanglement by studying the concurrence of TMEECSs. A quantification of quantum correlations of quasi-Werner states, prepared by two superposed $m$-photon-added bipartite coherent states, has been done recently \cite{arpi}. In these previous literatures in the direction of ECSs, quantum entanglement is explored by adding single or equal number of photons to a bipartite superposition of coherent states with opposite phases in the form $\ket{\psi}\propto a^{\dagger m}\otimes b^{\dagger m}(\ket{\alpha}_a\ket{\alpha}_b\pm
\ket{-\alpha}_a\ket{-\alpha}_b)$, where $a^\dagger (b^\dagger)$ is the photon creation operator for mode $a$ ($b$). In one of the introductory article, the authors explored the entanglement property of TMEECSs with a different number of excitations by means of concurrence. But while observing the influence of the photon excitations on the quantum entanglement of the TMEECS $\ket{\psi_{\pm}(\alpha, m, n)}$, they considered the specific case $m = n$ in which there are the same photon excitations in each filed modes
of the TMEECS. However, a full characterization including a graphical illustration of the entanglement properties in both pure and mixed systems is still needed to evaluate the effect of an arbitrary superposition with different coherent states and different photon addition numbers. We study (both analytically and graphically) the impact of two control parameters, coherent state amplitude $|\alpha|$ and photon-excitation pair $(m, n)$, over a few entanglement measures, namely $P$ function, $Q$ representation and von Neumann entropy.

In the present work, we describe a class of continuous-variable entangled states on the basis of entangled bipartite
coherent states (ECSs) \cite{mun}, called the photon-added entangled coherent states (PAECSs), which are obtained
by the actions of creation operator on ECSs. We investigate the entanglement characteristics of the PAECSs by analyzing the Schmidt decomposition and entropy of entanglement. The paper is structured as follows. In Sect.\,(\ref{sec2}), we present the definition of the PAECSs we are considering here and write down their forms in
terms of Fock states and some results for their scalar products. In Sect.\,(\ref{sec3}), we investigate the Schmidt decomposition in terms of the excited even (odd) coherent state. Then we calculate the von Neumann entropy for the PAECSs and discuss the influence of different excitation photon numbers on the entropy of entanglement. The last section ends with a summary of the main results of this article.

%On the other hand, a class of nonclassical states generated by applying creation operator on certain states is also an area of interest. The
%photon-added coherent states (PACS) \cite{agar} or excited coherent states \cite{nath}, first introduced by Agarwal and Tara, are the result
%of successive elementary one-photon excitations of a coherent state. The PACSs nonclassical quasi-probability distribution
%and amplitude are discussed in detail \cite{hu}. The groups of Zavatta \cite{zavatta} and Kalamidas \cite{kala} prepared experimentally a
%single PACS and a two PACS by using parametric down-conversion, respectively. Similarly, photon-added squeezed states
%\cite{zhang}, photon-added even and (or) odd coherent states \cite{dodo}, modified photon-added coherent states \cite{liang}, photon-added thermal
%states \cite{jones}, and so on, have already presented.

\section{Photon-added entangled coherent states}
\label{sec2}
This section begins with entangled coherent states (ECSs) defined as \cite{mun,gerry1}
\begin{eqnarray}
\left.
\begin{array}{rcl}
\label{eq1}
\ket{\psi_1^\pm(\alpha,0,0)} & = & N_{00}^\pm\big(\ket{\alpha,\alpha}\pm\ket{-\alpha,-\alpha}\big)\\\\
\ket{\psi_2^\pm(\alpha,0,0)} & = & N_{00}^\pm\big(\ket{\alpha,-\alpha}\pm\ket{-\alpha,\alpha}\big)
\end{array}
\right\}
\end{eqnarray}
where $\ket{\alpha,\alpha}\equiv {\ket{\alpha}}_a\otimes{\ket{\alpha}}_b$ is a bipartite coherent state with $\ket\alpha$ being a usual coherent state, defined by applying the displacement operator $D(\alpha) = e^{\alpha a^\dagger-\alpha^* a}$ upon the vacuum state. By using the overlap $\braket{\alpha|-\alpha}=e^{-2|\alpha|^2}$, the normalization constants can be calculated as
\begin{eqnarray}
N_{00}^\pm = \Big\{2\big(1\pm e^{-4|\alpha|^2}\big)\Big\}^{-1/2}.
\end{eqnarray}
Due to the overcompleteness of coherent states, $\bra\alpha-\alpha\rangle \neq 0$ for finite values of $\alpha$, thus the
ECSs are not mutually orthogonal. However, the overlap $\bra\alpha-\alpha\rangle$ tends to zero very rapidly with increase of $\alpha$. Now, our states of interest (PAECSs) are $m$-photon excitations of the mode $a$ and $n$-photon excitations of the mode $b$ on the ECSs, respectively, which are expressed as
\begin{eqnarray}
\left.
\begin{array}{rcl}
\label{eq2}
\ket{\psi_1^\pm(\alpha,m,n)} & = & N_{mn}^\pm a^{\dagger m}b^{\dagger n}\big(\ket{\alpha,\alpha}\pm\ket{-\alpha,-\alpha}\big)\\\\
\ket{\psi_2^\pm(\alpha,m,n)} & = & N_{mn}^\pm a^{\dagger m}b^{\dagger n}\big(\ket{\alpha,-\alpha}\pm\ket{-\alpha,\alpha}\big)
\end{array}
\right\}
\end{eqnarray}
To find out the normalization factor $N_{mn}^\pm$, we first derive an operator identity, the normal ordering form of the Boson operator $a^{\dagger m}b^{\dagger n}$. Using the completeness of the coherent state $\frac{1}{\pi}\int d^2\alpha \ket\alpha\bra\alpha = 1$ and the technique of integration within an ordered product of operators (IWOP) \cite{fan1} as well as the vacuum projector $\ket0\bra0=:\exp(-a^\dagger a):$ (:: represents the normal
ordering), we have
\begin{eqnarray}\nonumber
\label{eq4}
a^n a^{\dagger m} & = &\int \frac{d^2\alpha}{\pi} a^n\ket\alpha\bra\alpha a^{\dagger m}\\\nonumber
& = & \int \frac{d^2\alpha}{\pi} \alpha^n \alpha^{* m}:\exp(-|\alpha|^2+\alpha a^\dagger+\alpha^* a-a^\dagger a):\\
& = & (-i)^{m+n}:H_{m,n}(ia^\dagger, ia):
\end{eqnarray}
where
\begin{eqnarray*}
H_{m,n}(\xi, \eta) = (-1)^n e^{\xi\eta}\int \frac{d^2z}{\pi} z^n z^{* m}:\exp(-|z|^2+\xi z+\xi z^*):
\end{eqnarray*}
is the integral form of the two-variable Hermite polynomials \cite{fan2} and the two-variable Hermite polynomial is given by
\begin{eqnarray}
H_{m,n}(\xi, \eta) = \sum_{l=0}^{\mathrm{min}(m, n)} \frac{(-1)^{l}m!n!}{l!(m-l)!(n-l)!}\xi^{m-l}\eta^{n-l}.
\end{eqnarray}
Using (\ref{eq4}), the expectation of $a^m a^{\dagger m}$ with respect to the coherent state $\ket\alpha$ can be calculated as
\begin{eqnarray}\nonumber
\label{eq6}
\bra\alpha a^m a^{\dagger m} \ket\alpha & = & (-1)^m H_{m, m}(i\alpha^*, i\alpha)\\
& = & m!\,\, L_m(-|\alpha|^2)
\end{eqnarray}
where $L_m(x)$ is the $m$-th order Laguerre polynomial and is defined as
\begin{eqnarray}
L_m(x) & = & \sum_{l=0}^{m} \frac{(-1)^{l}m!}{(l!)^2(m-l)!} x^l.
\end{eqnarray}
In a similar way, using (\ref{eq6}) we obtain
\begin{eqnarray}\nonumber
\label{eq8}
\bra\alpha a^m a^{\dagger m} \ket{-\alpha} & = & (-1)^m e^{-2|\alpha|^2}H_{m,n}(i\alpha^*, -i\alpha)\\
& = & m!\,e^{-2|\alpha|^2}\,\, L_m(|\alpha|^2)
\end{eqnarray}
Using the results in (\ref{eq6}) and (\ref{eq8}), the normalization constant $N_{mn}^\pm$ can be obtained as
\begin{eqnarray}\nonumber
N_{mn}^\pm
& = & \Big[2\,m!\,n!\,\big\{L_m(|-\alpha|^2)L_n(-|\alpha|^2)\\\nonumber
& & \pm e^{-4|\alpha|^2}L_m(|\alpha|^2)L_n(|\alpha|^2)\big\}\Big]^{-1/2}\\\nonumber
& = & \Big[2\big\{L_{m, n}(|-\alpha|^2, -|\alpha|^2)\\
& & \pm e^{-4|\alpha|^2}L_{m, n}(|\alpha|^2, |\alpha|^2)\big\}\Big]^{-1/2}
\end{eqnarray}
by introducing the notation $L_{m, n}(x, y)=m!\,n!\,L_m(x)\,L_n(y)$.
Specially, in the limit $m\rightarrow 0$ and $n\rightarrow0$, $\ket{\psi_1^\pm(\alpha,m,n)}$ and $\ket{\psi_2^\pm(\alpha,m,n)}$ reduce to the usual entangled coherent states in (\ref{eq1}). If $m=0$ or $n=0$, the excitations turns to single-mode excited coherent states \cite{xu}. If $m=n\neq 0$, the excitations turn to two-mode excited entangled coherent states \cite{zhang}. So studying $\ket{\psi_1^\pm(\alpha,m,n)}$ and $\ket{\psi_2^\pm(\alpha,m,n)}$ will help to conclude about the behavior of all these states. Using $a^{\dagger m}\ket n=\sqrt{\frac{(n+m)!}{n!}}\ket {n+m}$ and the Fock state representation of a coherent state, we have obtained the following expressions

\begin{eqnarray*}
a^{\dagger m}b^{\dagger n}\ket{\alpha, \alpha} & = & e^{-|\alpha|^2}\sum_{p, q=0}^\infty \frac{\sqrt{(p+m)!(q+n)!}}{p!q!}\\
& & {\alpha}^{p+q}\ket{p+m, q+n}\\
a^{\dagger m}b^{\dagger n}\ket{-\alpha, -\alpha} & = & e^{-|\alpha|^2}\sum_{p, q=0}^\infty \frac{\sqrt{(p+m)!(q+n)!}}{p!q!}\\
& & (-\alpha)^{p+q}\ket{p+m, q+n}\\
a^{\dagger m}b^{\dagger n}\ket{\alpha, -\alpha} & = & e^{-|\alpha|^2}\sum_{p, q=0}^\infty \frac{\sqrt{(p+m)!(q+n)!}}{p!q!}\\
& & (-1)^q{\alpha}^{p+q}\ket{p+m, q+n}\\
a^{\dagger m}b^{\dagger n}\ket{-\alpha, \alpha} & = & e^{-|\alpha|^2}\sum_{p, q=0}^\infty \frac{\sqrt{(p+m)!(q+n)!}}{p!q!}\\
& & (-1)^p{\alpha}^{p+q}\ket{p+m, q+n}
\end{eqnarray*}
Therefore, (\ref{eq2}) can be rewritten as
\begin{widetext}
\begin{eqnarray*}
\ket{\psi_1^\pm(\alpha,m,n)} & = & N_{mn}^\pm e^{-|\alpha|^2}\sum_{p, q=0}^\infty \frac{\sqrt{(p+m)!(q+n)!}}{p!q!} {\alpha}^{p+q}\big[1\pm(-1)^{p+q}\big]\ket{p+m, q+n}\\
\ket{\psi_2^\pm(\alpha,m,n)} & = & N_{mn}^\pm e^{-|\alpha|^2}\sum_{p, q=0}^\infty \frac{\sqrt{(p+m)!(q+n)!}}{p!q!} {\alpha}^{p+q}\big[(-1)^q\pm(-1)^p\big]\ket{p+m, q+n}
\end{eqnarray*}
\end{widetext}
which implies that they are actually truncations of the two-mode ECSs given by an equation with respect to the mode $a$, $b$ where
all the terms related to the Fock states of the mode $a: \ket{0}, \ket{1},\ldots,\ket{m-1}$ and the Fock states of the mode $b: \ket{0}, \ket{1},\ldots,\ket{n-1}$ are removed. Finally, this expansion also leads to the following results for the scalar products:
\begin{widetext}
\begin{eqnarray}\nonumber
\langle\psi_1^\pm(\alpha,m',n')|\psi_1^\pm(\beta,m,n)\rangle & = & N_{m'm}^\pm N_{n'n}^\pm\Big[A_{m'm}(\alpha, \beta)A_{n'n}(\alpha, \beta)+A_{m'm}(-\alpha, -\beta)A_{n'n}(-\alpha, -\beta)\\
& &
\pm A_{m'm}(\alpha, -\beta)A_{n'n}(\alpha, -\beta)\pm A_{m'm}(-\alpha, \beta)A_{n'n}(-\alpha, \beta)\Big]
\end{eqnarray}
\begin{eqnarray}\nonumber
\langle\psi_2^\pm(\alpha,m',n')|\psi_2^\pm(\beta,m,n)\rangle & = & N_{m'm}^\pm N_{n'n}^\pm\Big[A_{m'm}(\alpha, \beta)A_{n'n}(-\alpha, -\beta)+A_{m'm}(-\alpha, -\beta)A_{n'n}(\alpha, \beta)\\
& &
\pm A_{m'm}(\alpha, -\beta)A_{n'n}(-\alpha, \beta)\pm A_{m'm}(-\alpha, \beta)A_{n'n}(\alpha, -\beta)\Big]
\end{eqnarray}
and
\begin{eqnarray}\nonumber
\langle\psi_1^\pm(\alpha,m',n')|\psi_2^\pm(\beta,m,n)\rangle & = & N_{m'm}^\pm N_{n'n}^\pm\Big[A_{m'm}(\alpha, \beta)A_{n'n}(\alpha, -\beta)+A_{m'm}(-\alpha, -\beta)A_{n'n}(-\alpha, \beta)\\
& &
\pm A_{m'm}(\alpha, -\beta)A_{n'n}(\alpha, \beta)\pm A_{m'm}(-\alpha, \beta)A_{n'n}(-\alpha, -\beta)\Big]
\end{eqnarray}
\end{widetext}
where
\begin{eqnarray}\nonumber
A_{mn}(\alpha, \beta) & = & (-i)^{m+n}\exp\left\{-\frac{|\alpha|^2}{2}-\frac{|\beta|^2}{2}+\alpha^*\beta\right\}\\
& & H_{m,n}(i\alpha^*,i\beta).
\end{eqnarray}

\section{Entanglement of photon-added entangled coherent states}
\label{sec3}
Let the normalized even and odd coherent states be
\begin{eqnarray}
\ket{\alpha_{\mathrm{e}}}=\frac{1}{\sqrt{2(1+e^{-2|\alpha|^2})}}\Big[\ket\alpha+\ket{-\alpha}\Big]
\end{eqnarray}
and
\begin{eqnarray}
\ket{\alpha_{\mathrm{o}}}=\frac{1}{\sqrt{2(1-e^{-2|\alpha|^2})}}\Big[\ket\alpha-\ket{-\alpha}\Big]
\end{eqnarray}
Therefore,
%\begin{widetext}
\begin{eqnarray}
\label{eq16}
\left.
\begin{array}{rcl}
\ket{\alpha, \alpha}+\ket{-\alpha, -\alpha} & = & (1+e^{-2|\alpha|^2})\ket{\alpha_{\mathrm{e}}}\ket{\alpha_{\mathrm{e}}}\\
& & +(1-e^{-2|\alpha|^2})\ket{\alpha_{\mathrm{o}}}\ket{\alpha_{\mathrm{o}}}\\\\
\ket{\alpha, \alpha}-\ket{-\alpha, -\alpha} & = & \sqrt{1-e^{-4|\alpha|^2}}\Big[\ket{\alpha_{\mathrm{o}}}\ket{\alpha_{\mathrm{e}}}\\
& & +\ket{\alpha_{\mathrm{e}}}\ket{\alpha_{\mathrm{o}}}\Big]\\\\
\ket{\alpha, -\alpha}+\ket{-\alpha, \alpha} & = & (1+e^{-2|\alpha|^2})\ket{\alpha_{\mathrm{e}}}\ket{\alpha_{\mathrm{e}}}\\
& & -(1-e^{-2|\alpha|^2})\ket{\alpha_{\mathrm{o}}}\ket{\alpha_{\mathrm{o}}}\\\\
\ket{\alpha, -\alpha}-\ket{-\alpha, \alpha} & = & \sqrt{1-e^{-4|\alpha|^2}}\Big[\ket{\alpha_{\mathrm{o}}}\ket{\alpha_{\mathrm{e}}}\\
& & -\ket{\alpha_{\mathrm{e}}}\ket{\alpha_{\mathrm{o}}}\Big]
\end{array}
\right\}
\end{eqnarray}
%\end{widetext}
By substituting (\ref{eq16}) into (\ref{eq2}), we have obtained
%\begin{widetext}
\begin{eqnarray}\nonumber
\label{eq17}
\ket{\psi_1^+(\alpha,m,n)} & = & N_{mn}^+ e^{-|\alpha|^2}\Big[\sqrt{L_{m+,n+}}\ket{\alpha_{\mathrm{e}},m}\ket{\alpha_{\mathrm{e}},n}\\
& & + \sqrt{L_{m-,n-}}\ket{\alpha_{\mathrm{o}},m}\ket{\alpha_{\mathrm{o}},n}\Big]
\end{eqnarray}
\begin{eqnarray}\nonumber
\label{eq18}
\ket{\psi_1^-(\alpha,m,n)} & = & N_{mn}^- e^{-|\alpha|^2}\Big[\sqrt{L_{m-,n+}}\ket{\alpha_{\mathrm{o}},m}\ket{\alpha_{\mathrm{e}},n}\\
& & + \sqrt{L_{m+,n-}}\ket{\alpha_{\mathrm{e}},m}\ket{\alpha_{\mathrm{o}},n}\Big]
\end{eqnarray}
\begin{eqnarray}\nonumber
\label{eq19}
\ket{\psi_2^+(\alpha,m,n)} & = & N_{mn}^+ e^{-|\alpha|^2}\Big[\sqrt{L_{m+,n+}}\ket{\alpha_{\mathrm{e}},m}\ket{\alpha_{\mathrm{e}},n}\\
& & - \sqrt{L_{m-,n-}}\ket{\alpha_{\mathrm{o}},m}\ket{\alpha_{\mathrm{o}},n}\Big]
\end{eqnarray}
\begin{eqnarray}\nonumber
\label{eq20}
\ket{\psi_2^-(\alpha,m,n)} & = & N_{mn}^- e^{-|\alpha|^2}\Big[\sqrt{L_{m-,n+}}\ket{\alpha_{\mathrm{o}},m}\ket{\alpha_{\mathrm{e}},n}\\
& & - \sqrt{L_{m+,n-}}\ket{\alpha_{\mathrm{e}},m}\ket{\alpha_{\mathrm{o}},n}\Big]
\end{eqnarray}
%\end{widetext}
where $\ket{\alpha_{\mathrm{e}},m}$ ($\ket{\alpha_{\mathrm{o}},n}$) is the normalized even (odd) coherent state \cite{xin}, $$L_m^\pm(x)=e^x L_m(-x)\pm e^{-x} L_m(x)$$ and $$L_{m\pm, n\pm}=m!\,n!\,L_m^\pm(|\alpha|^2)\,L_n^\pm(|\alpha|^2)$$ Equations (\ref{eq17})-(\ref{eq20}) give the Schmidt decomposition in terms of the excited even (odd) coherent states. It can be seen that when $m=n$, $\ket{\psi_1^-(\alpha,m,n)}$ and $\ket{\psi_2^-(\alpha,m,n)}$ are always the maximally entangled states.

Here we use the entropy of entanglement to analyze the entanglement properties of PAECSs. The von Neumann entropy \cite{bennet} for the reduced density matrix $\rho_b\equiv \mathrm{Tr}_a[\ket\varphi_{ab}\bra\varphi_{ab}]$ of a pure entangled state $\ket\varphi_{ab}$ is defined as $E(\ket\varphi_{ab}) = -\mathrm{Tr}[\rho_b \log_2\rho_b]$. So by applying Schmidt decomposition of PAECSs in (\ref{eq17})-(\ref{eq20}), the reduced density matrix $\rho_b$ for $\ket{\psi_1^+(\alpha,m,n)}$ can be expressed as
\begin{eqnarray}\nonumber
\rho_b & = & \mathrm{Tr}_a[\ket{\psi_1^+(\alpha,m,n)}\bra{\psi_1^+(\alpha,m,n)}]\\\nonumber
& = & N_{mn}^+ e^{-2|\alpha|^2}\Big(L_{m+,n+}\ket{\alpha_{\mathrm{e}},n}\bra{\alpha_{\mathrm{e}},n}\\
& & +L_{m-,n-}\ket{\alpha_{\mathrm{o}},n}\bra{\alpha_{\mathrm{o}},n} \Big)
\end{eqnarray}
Now, the two nonnegative eigenvalues $\lambda_\pm(\psi_1^+)$ of $\rho_b$ are as follows:
\begin{eqnarray}\nonumber
\lambda_\pm(\psi_1^+) = \frac{1}{2}\pm\sqrt{\frac{1}{4}-(N_{mn}^+)^2 e^{-4|\alpha|^2}L_{m+,n+}L_{m-,n-}}\\
\label{eq22}
\end{eqnarray}
Similarly, the two nonnegative eigenvalues of other PAECSs are
\begin{eqnarray}\nonumber
\lambda_\pm(\psi_1^-) = \frac{1}{2}\pm\sqrt{\frac{1}{4}-(N_{mn}^-)^2 e^{-4|\alpha|^2}L_{m+,n-}L_{m-,n+}}\\
\label{eq23}
\end{eqnarray}
\begin{eqnarray}\nonumber
\lambda_\pm(\psi_2^+) = \frac{1}{2}\pm\sqrt{\frac{1}{4}-(N_{mn}^+)^2 e^{-4|\alpha|^2}L_{m+,n+}L_{m-,n-}}\\
\label{eq24}
\end{eqnarray}
and
\begin{eqnarray}\nonumber
\lambda_\pm(\psi_2^-) = \frac{1}{2}\pm\sqrt{\frac{1}{4}-(N_{mn}^-)^2 e^{-4|\alpha|^2}L_{m+,n-}L_{m-,n+}}\\
\label{eq25}
\end{eqnarray}
respectively. Thus, the entropy of entanglement of PAECSs are
\begin{eqnarray}
E(\ket{\psi}) = -\lambda_+ \log_2 \lambda_+ -\lambda_- \log_2 \lambda_-.
\end{eqnarray}
It is interesting to note that the entropy of $\ket{\psi_1^+(\alpha, m, n)}$ is equal to the entropy of $\ket{\psi_2^+(\alpha, m, n)}$ and the entropy of $\ket{\psi_1^-(\alpha, m, n)}$ is equal to the entropy of $\ket{\psi_2^-(\alpha, m, n)}$.

\begin{figure}[htbp]
\centering
\includegraphics[scale=0.5]{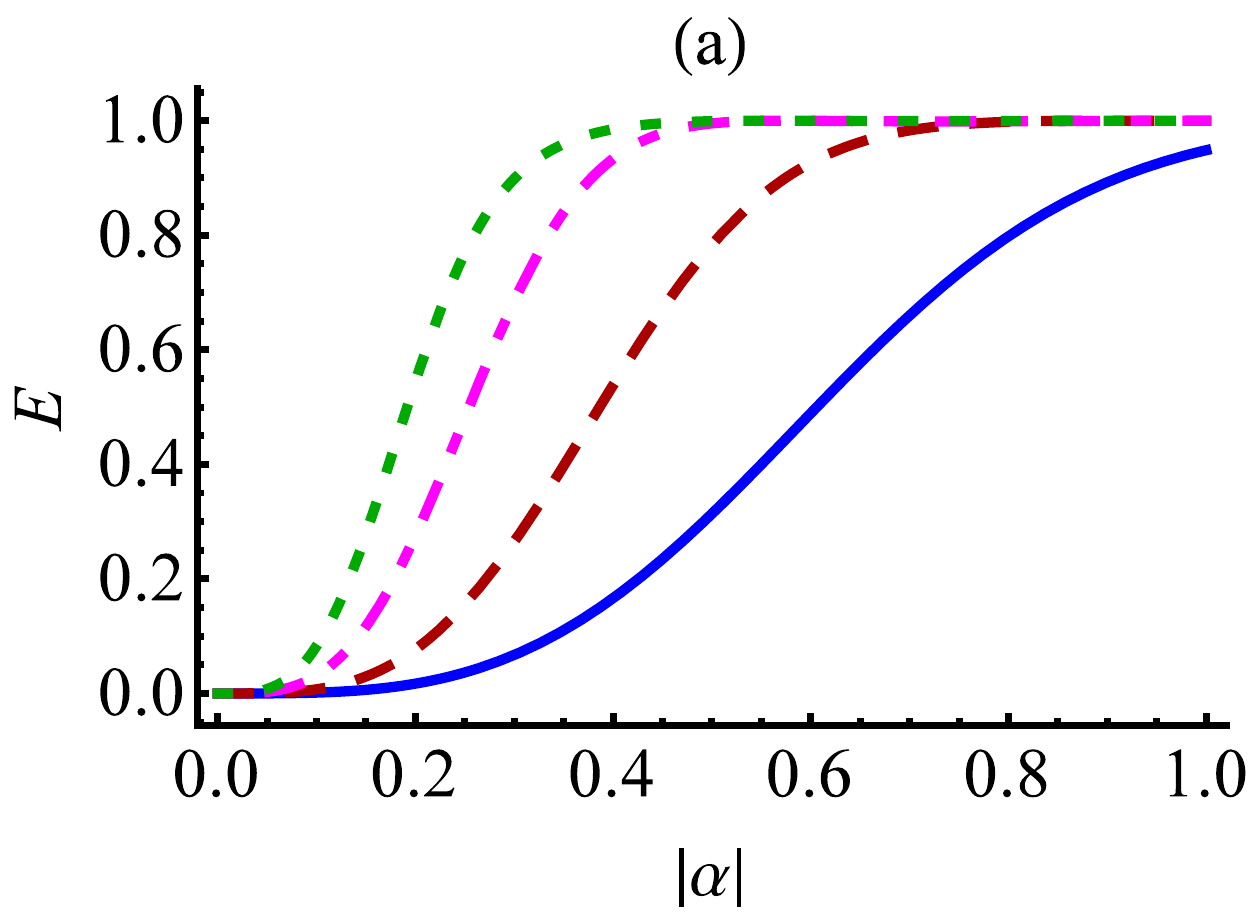}
\includegraphics[scale=0.5]{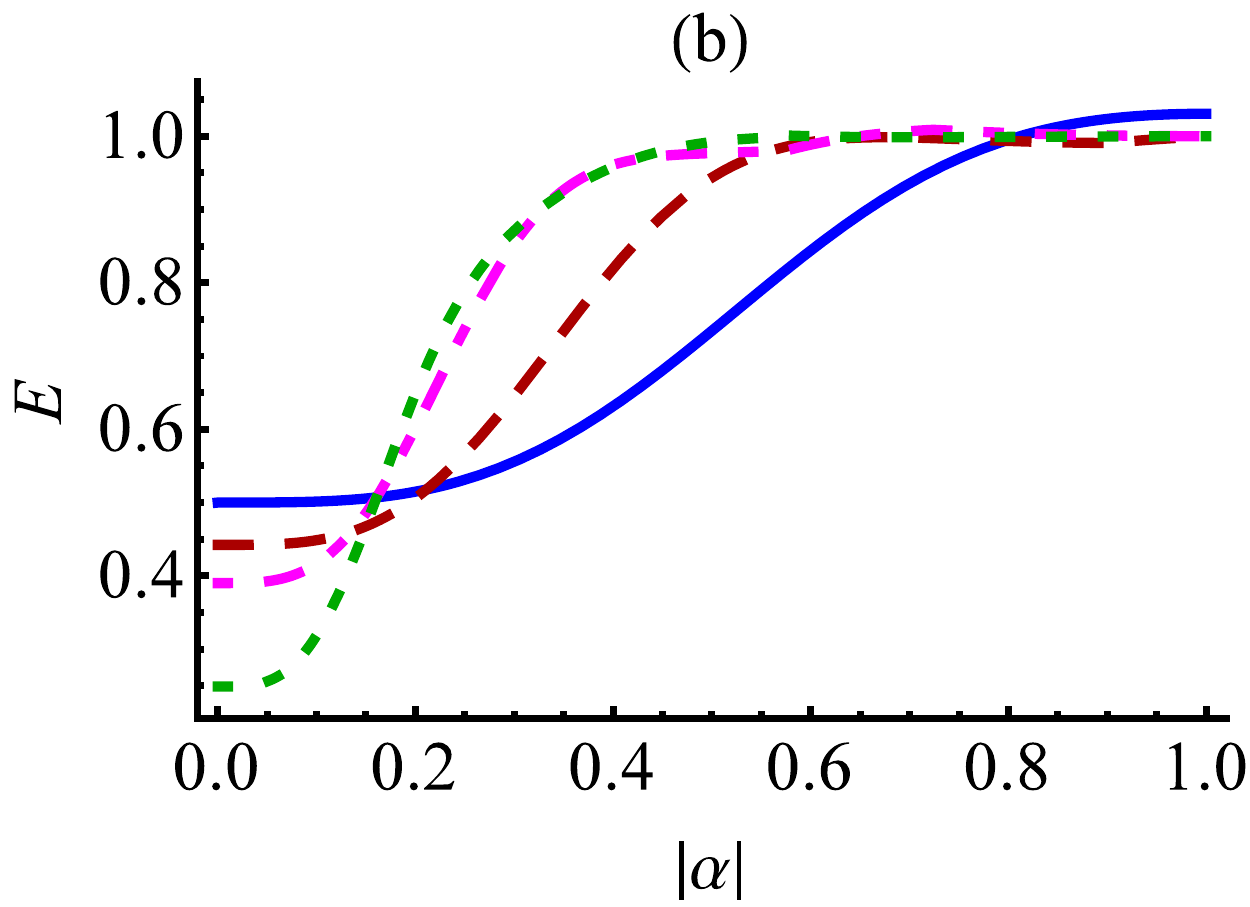}

\caption{(Color online) The entropy of entanglement for the states (a) $\ket{\psi_1^+(\alpha, m, n)}$ and (b) $\ket{\psi_1^-(\alpha, m, n)}$ as a function of $|\alpha|$
for $m=n=0$ (solid blue line), $m=2$, $n=1$ (darker red dashed line), $m=3$, $n=7$ (magenta dash-dotted line) and $m=20$, $n=4$ (green dotted line).}
\label{fig1}
\end{figure}

\begin{figure}[htbp]
\centering
\includegraphics[scale=0.5]{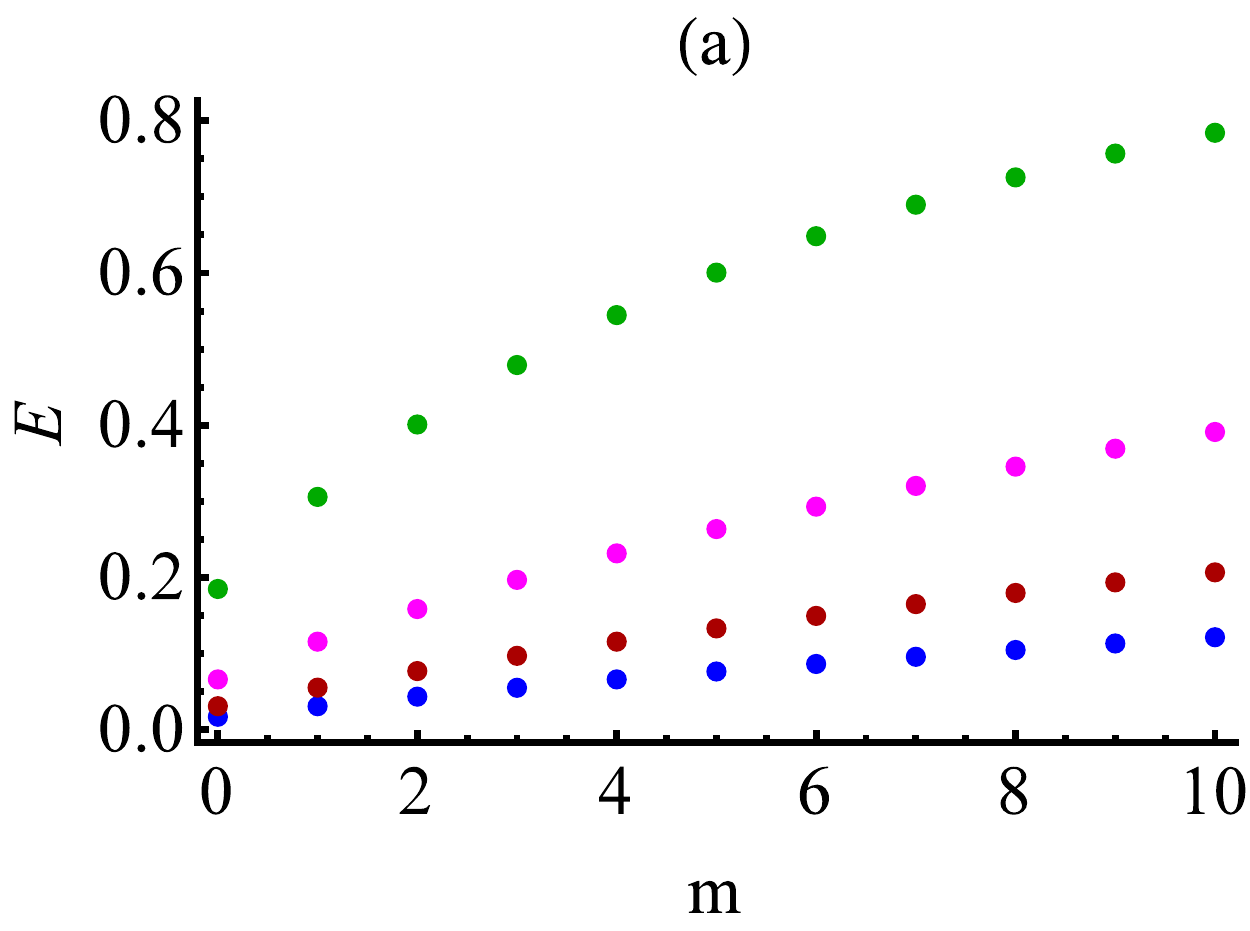}
\includegraphics[scale=0.5]{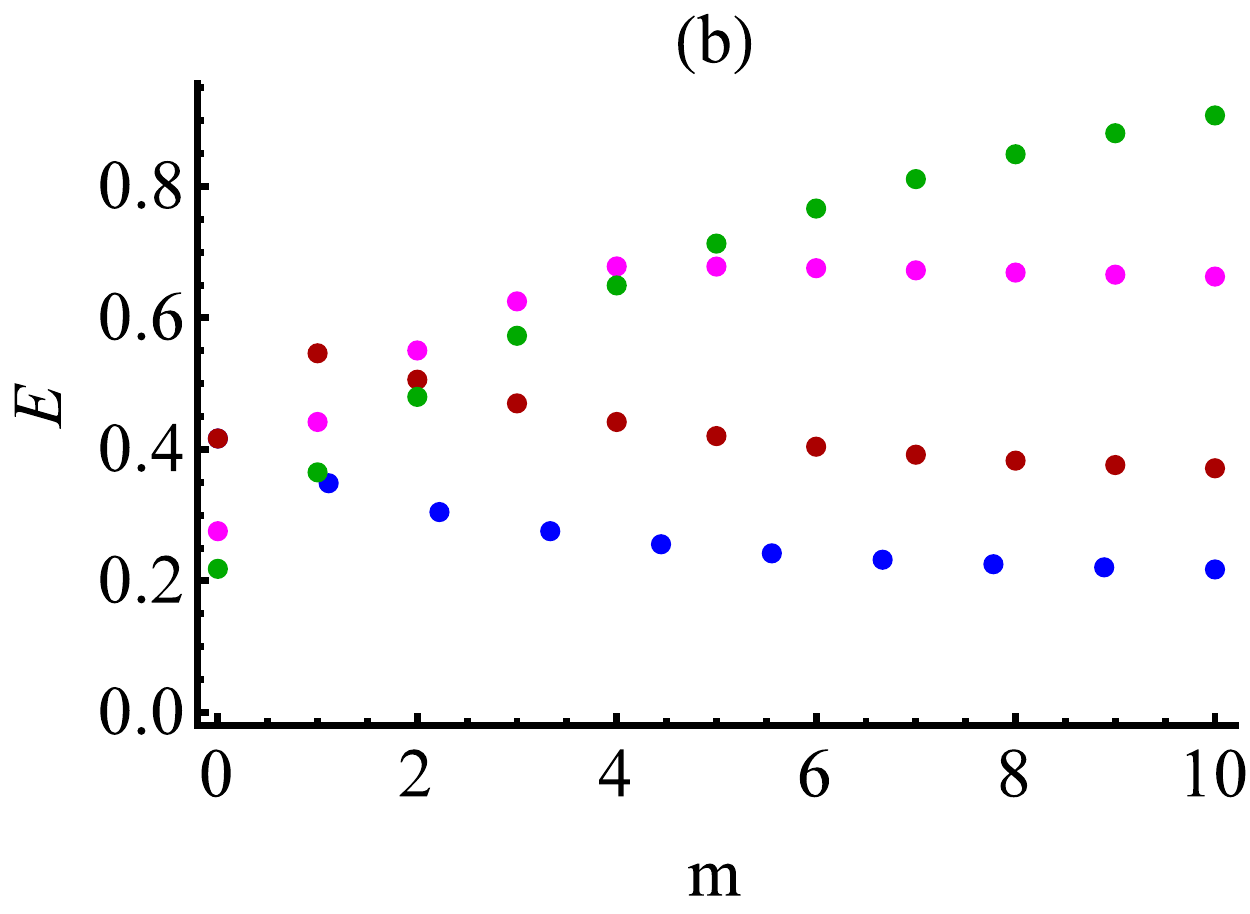}

\caption{(Color online) The entropy of entanglement for the states (a) $\ket{\psi_1^+(\alpha, m, n)}$ and (b) $\ket{\psi_1^-(\alpha, m, n)}$ as a function of $m$ and for $n=0$ (blue points), $n=1$ (darker red points), $n=4$ (magenta points) and $n=20$ (green points) with $|\alpha|=0.2$.}
\label{fig2}
\end{figure}
In order to detect the influence of the photon excitations on the quantum entanglement of PAECSs, we plot $E$ as a function of ${|\alpha|}$ in
Fig.~\ref{fig1} for different values of $m$ and $n$. We find that interchanging the number of photon excitations between two
modes does merely affect the entanglement amount of PAECSs, that means $E(\alpha, m, n) = E(\alpha, n, m)$. One can see that $E$ is sensitive to the values
of photon excitations $m$, $n$ and increases while $m$, $n$ are increasing from 2 to 20 and 1 to 4, respectively. When $m=n=0$, our result is in accordance with \cite{serna} where Francisco et al. observed that concurrence $C(\psi^{\mathrm{ AB}})$ increases asymptotically from 0 to 1 as $\alpha$ increases. Thus photon addition process can be identified as an entanglement enhancer operation for superpositions of coherent states (SCS). In Fig.~\ref{fig1}(b), $E$ of $\ket{\psi_1^-(\alpha, m, n)}$ against ${|\alpha|}$ is shown. It is clear from (\ref{eq22})-(\ref{eq25}) that the behaviour of $\ket{\psi_2^+(\alpha, m, n)}$ ($\ket{\psi_2^-(\alpha, m, n)}$) is similar as $\ket{\psi_1^+(\alpha, m, n)}$ ($\ket{\psi_1^-(\alpha, m, n)}$). As $m = n$, $E(\ket{\psi_1^-})$ and $E(\ket{\psi_2^-})$ reach their maximal value 1, which implies that $\ket{\psi_1^-(\alpha, m, n)}$ and $\ket{\psi_2^-(\alpha, m, n)}$ are always maximally entangled states. Specifically, in the region of very small $|\alpha|$, the entropy of $\ket{\psi_1^+(\alpha, m, n)}$ does not depend much more on photon excitation pair ($m, n$). When $|\alpha|<0.2$, the most entangled $\ket{\psi_1^-(\alpha, m, n)}$ corresponds to the case $m=n=0$. This is because even for no photon excitations, there is an entangled coherent photon state. This makes the entangled photon-added coherent states more robust even for a small photon number. In Fig.~\ref{fig2}, the dependence of entropy on photon excitation number $m$ is presented for a given fixed $n$ and $|\alpha|=0.2$. It is shown that the entropy of entanglement for $\ket{\psi_1^+(\alpha, m, n)}$ increases steadily while $n$ is increasing from 0 to 20, keeping $|\alpha|$ fixed at 0.2. However, it is observed that the entropy of $\ket{\psi_1^-(\alpha, m, n)}$ (assuming $n=0,\,1$ or $4$), attains its maximal value when $m=0,\,1$ or $4$, respectively. That suggests in the case of $\ket{\psi_1^-(\alpha, m, n)}$, the state is most entangled when $m=n$.
%Here we can easily see that the entropy of entanglement of the PAECSs increases with ${|\alpha|}^2$.
%Especially, the entropy $E$ tends to unity for the larger ${|\alpha|}^2$. Therefore we can conclude that operating a creation
%operator can increase the amount of entanglement.

\section{$P$ and $Q$ functions of photon-added entangled coherent states}
The Glauber-Sudarshan $P$ distribution function gives a quasi-probability distribution in phase space, which can assume negative and singular values for nonclassical fields. If the Glauber-Sudarshan $P$-function exhibits a nonclassical character, then the state is entangled \cite{agar1}. In this section, we study the nonclassicality of the PAECSs characterized by the $P$-function and $Q$-function.
For a bipartite system, the density matrix can be expressed in terms of the diagonal two-mode coherent state $\ket{z_1, z_2}$ as \cite{glauber}
\begin{eqnarray}
\rho = \int {d^2 z_1 \, d^2 z_2 \, P(z_1, z_2) \ket{z_1, z_2}\bra{z_1, z_2} }
\end{eqnarray}
To calculate the $P$-function, we recall the antinormal ordering form of an arbitrary two-mode state $\rho$ by using \cite{fan3}
\begin{eqnarray}\nonumber
\label{eq28}
\rho & = & \int \frac{d^2z_1 \, d^2z_2}{\pi^2}\,\vdots\bra{-z_1, -z_2}\rho\ket{z_1, z_2}\\\nonumber
& & \exp\Big(|z_1|^2+|z_2|^2+{z_1}^* a-{z_1} a^\dagger+{z_2}^* b-{z_2} b^\dagger+a^\dagger a+b^\dagger b\Big)\vdots\\
\end{eqnarray}
where $\vdots \vdots$ stands for the antinormal ordering, $\ket{z_1, z_2}$ is a two-mode coherent state. Using $\braket{z_1| z_2}=\exp\left(-\frac{|z_1|^2}{2}-\frac{|z_2|^2}{2}+{z_1}^*z_2\right)$, the density matrix for the state $\ket{\psi_1^\pm(\alpha, m, n)}$ can be written as
\begin{eqnarray}\nonumber
\label{eq29}
\rho_{\psi_1^\pm} & = & \frac{(N_{mn}^\pm)^2 e^{-2|\alpha|^2}}{\pi^2}\vdots \exp(a^\dagger a+b^\dagger b)\frac{\partial^{2(m+n)}}{\partial \alpha^{*(m+n)} \partial \alpha^{(m+n)}}\\
& &\left[
                                                                      \begin{array}{c}
                                                                        \delta^2(\alpha-a) \delta^2(\alpha-b)\\
                                                                        \pm\delta(\alpha+a)\delta(\alpha^*-a^\dagger)\delta(\alpha^*-b^\dagger)\delta(\alpha+b)\\
                                                                        \pm\delta(\alpha-a)\delta(\alpha^*+a^\dagger)\delta(\alpha^*+b^\dagger)\delta(\alpha-b)\\
                                                                         \delta(\alpha+a)\delta(\alpha^*+a^\dagger)\delta(\alpha^*+b^\dagger)\delta(\alpha+b)
                                                                      \end{array}
                                                                    \right]\vdots
\end{eqnarray}
Similarly for $\ket{\psi_2^\pm(\alpha, m, n)}$,
\begin{eqnarray}\nonumber
\label{eq30}
\rho_{\psi_2^\pm} & = & \frac{(N_{mn}^\pm)^2 e^{-2|\alpha|^2}}{\pi^2}\vdots \exp(a^\dagger a+b^\dagger b)\frac{\partial^{2(m+n)}}{\partial \alpha^{*(m+n)} \partial \alpha^{(m+n)}}\\
& &\left[
                                                                      \begin{array}{c}
                                                                        \delta(\alpha+a)\delta(\alpha^*-a^\dagger)\delta(\alpha^*+b^\dagger)\delta(\alpha-b) \\
                                                                        \pm\delta^2(\alpha-a) \delta^2(\alpha+b)\\
                                                                        \pm\delta^2(\alpha+a) \delta^2(\alpha-b)\\
                                                                         \delta(\alpha-a)\delta(\alpha^*+a^\dagger)\delta(\alpha^*-b^\dagger)\delta(\alpha+b)
                                                                      \end{array}
                                                                    \right]\vdots
\end{eqnarray}
Comparing (\ref{eq29}) and (\ref{eq30}) with (\ref{eq28}), the $P$ functions can be obtained as
\begin{eqnarray}\nonumber
& & P_{\psi_1^\pm}(z_1, z_2)\\\nonumber & = & \frac{(N_{mn}^\pm)^2 e^{-2|\alpha|^2-|z_1|^2-|z_2|^2}}{\pi^2} \frac{\partial^{2(m+n)}}{\partial \alpha^{*(m+n)} \partial \alpha^{(m+n)}}\\
& &\left[
                                                                      \begin{array}{c}
                                                                        \delta^2(\alpha-z_1) \delta^2(\alpha-z_2)\\
                                                                        \pm\delta(\alpha+z_1)\delta(\alpha^*-z_1^*)\delta(\alpha^*-z_2^*)\delta(\alpha+z_2) \\
                                                                        \pm\delta(\alpha-z_1)\delta(\alpha^*+z_1^*)\delta(\alpha^*+z_2^*)\delta(\alpha-z_2) \\
                                                                         \delta(\alpha+z_1)\delta(\alpha^*+z_1^*)\delta(\alpha^*+z_2^*)\delta(\alpha+z_2)
                                                                      \end{array}
                                                                    \right]
\end{eqnarray}
and
\begin{eqnarray}\nonumber
& & P_{\psi_2^\pm}(z_1, z_2)\\\nonumber & = & \frac{(N_{mn}^\pm)^2 e^{-2|\alpha|^2-|z_1|^2-|z_2|^2}}{\pi^2} \frac{\partial^{2(m+n)}}{\partial \alpha^{*(m+n)} \partial \alpha^{(m+n)}}\\
& &\left[
                                                                      \begin{array}{c}
                                                                      \delta(\alpha+z_1)\delta(\alpha^*-z_1^*)\delta(\alpha^*+z_2^*)\delta(\alpha-z_2) \\
                                                                        \pm\delta^2(\alpha-z_1) \delta^2(\alpha+z_2)\\
                                                                        \pm\delta^2(\alpha+z_1) \delta^2(\alpha-z_2)\\
                                                                         \delta(\alpha-z_1)\delta(\alpha^*+z_1^*)\delta(\alpha^*-z_2^*)\delta(\alpha+z_2)
                                                                         \end{array}
                                                                    \right]
\end{eqnarray}

\begin{figure*}[htbp]
\centering
\includegraphics[width=5.7cm]{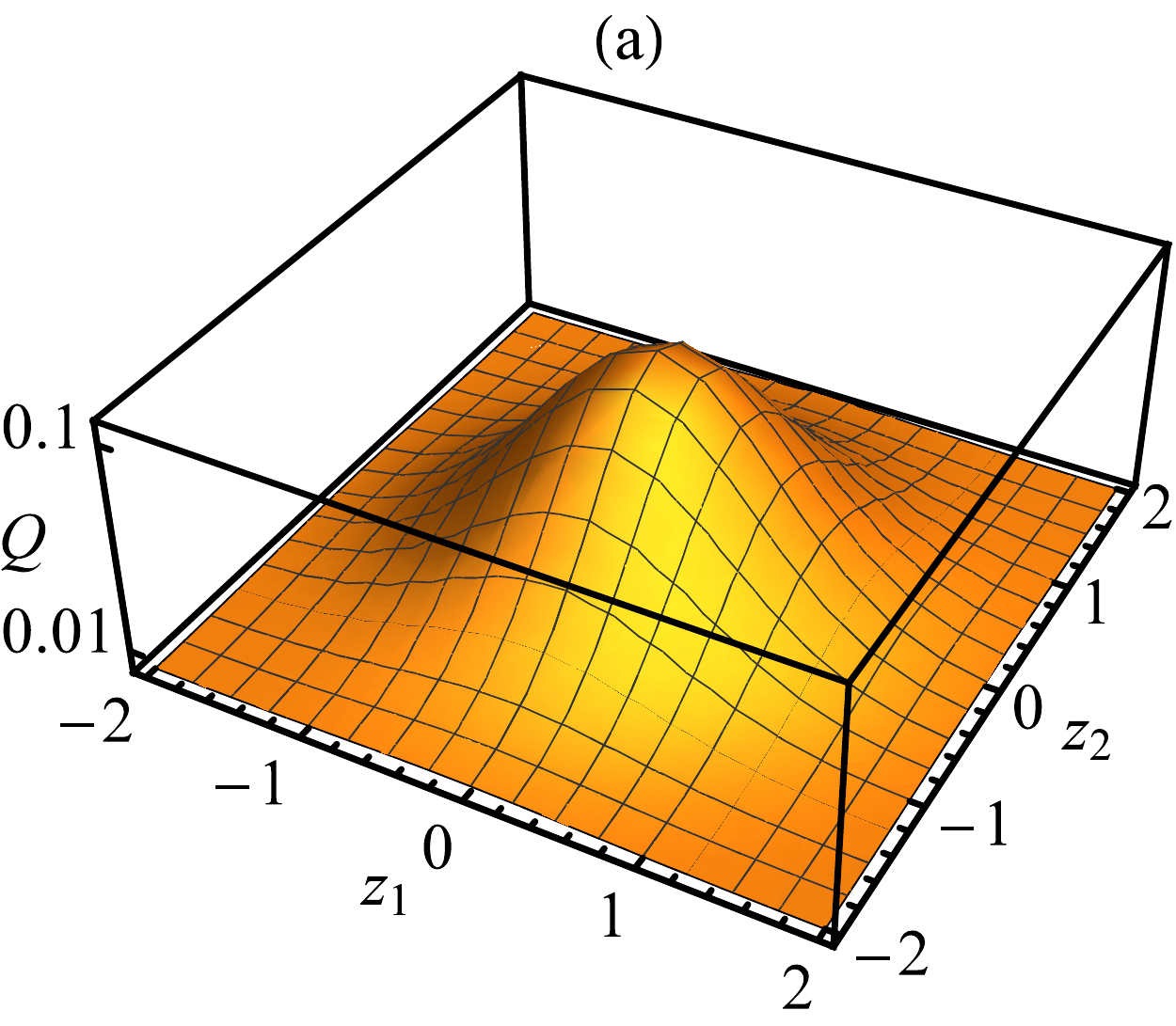}
\includegraphics[width=5.7cm]{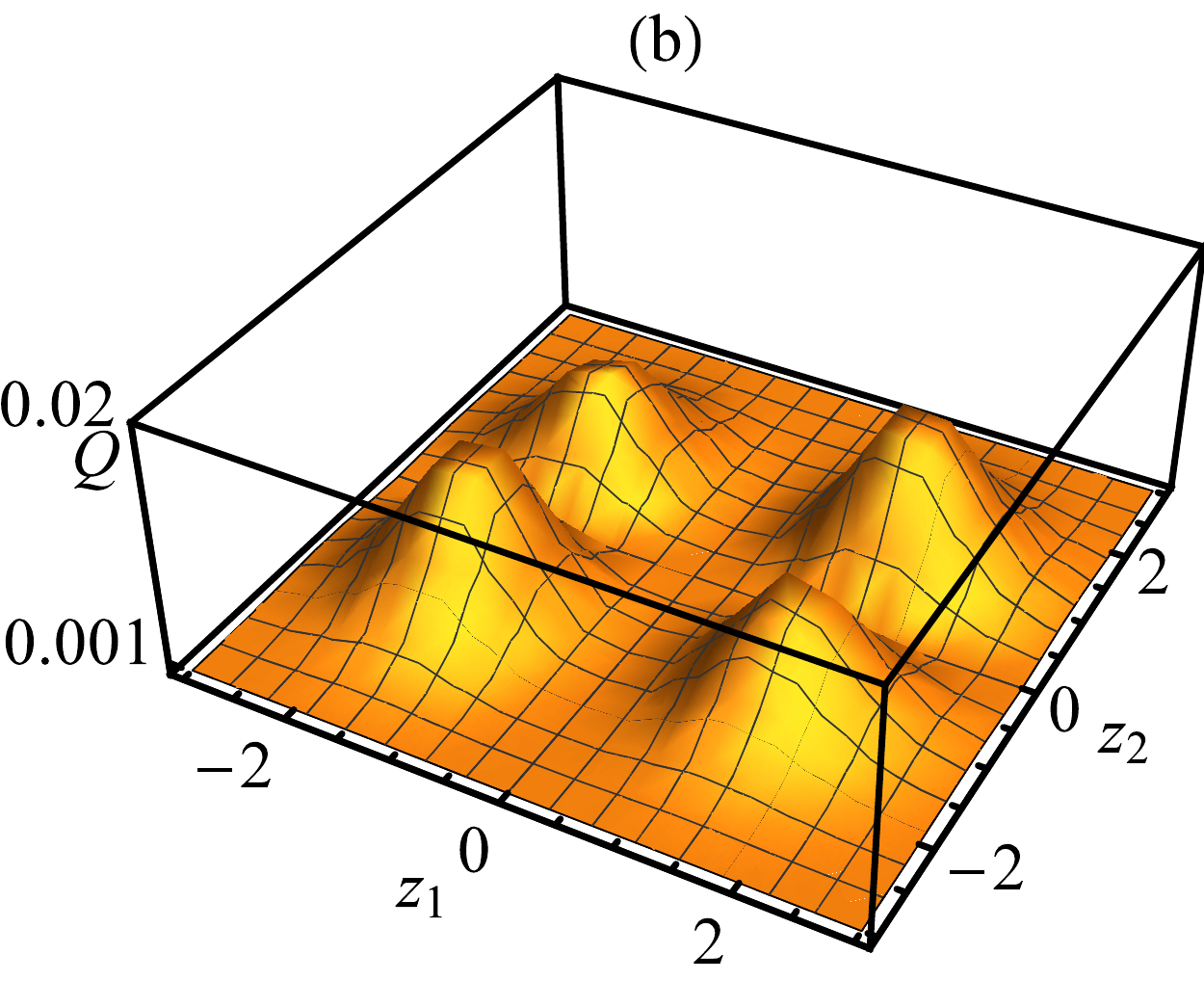}
\includegraphics[width=5.7cm]{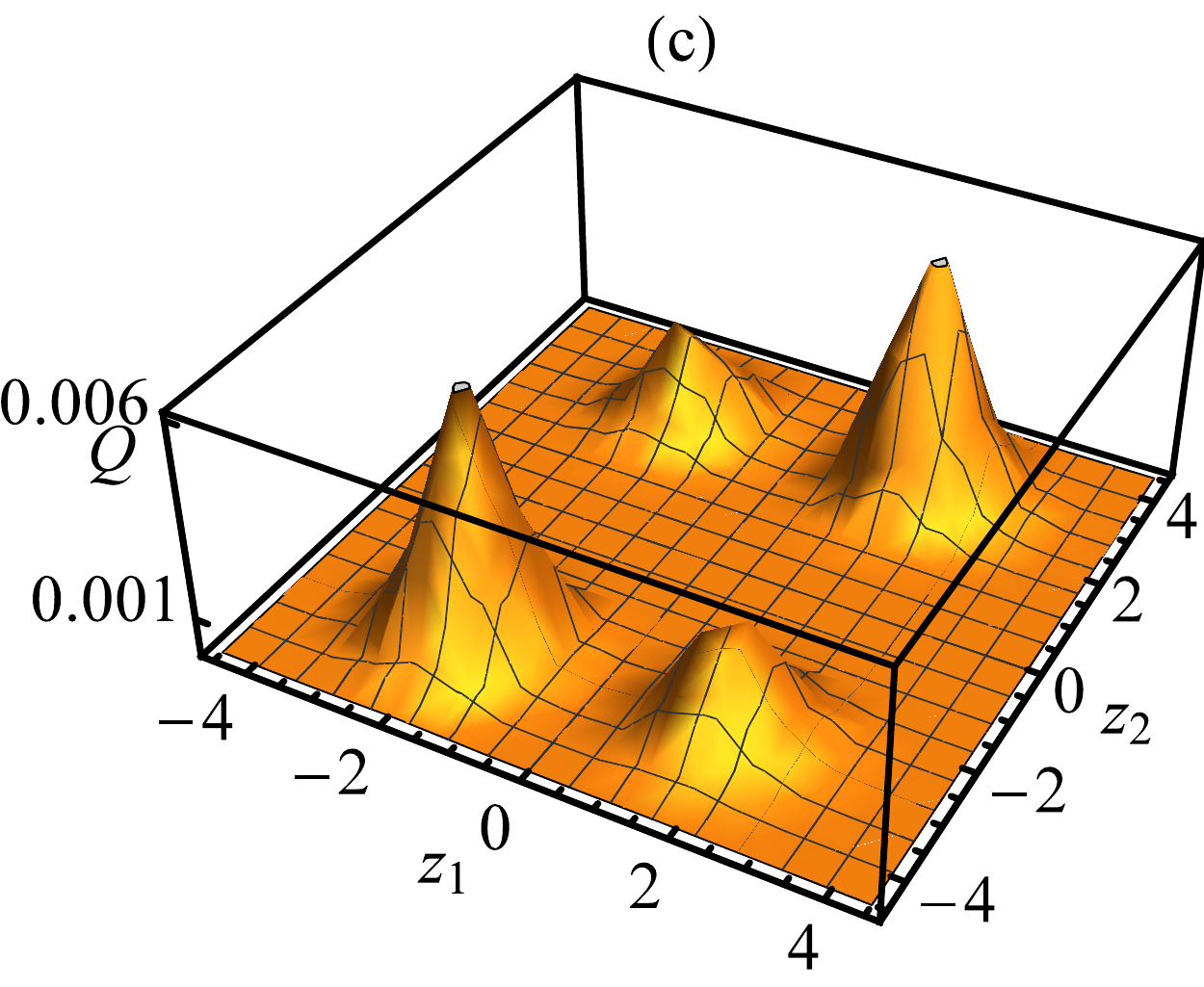}
\includegraphics[width=5.8cm]{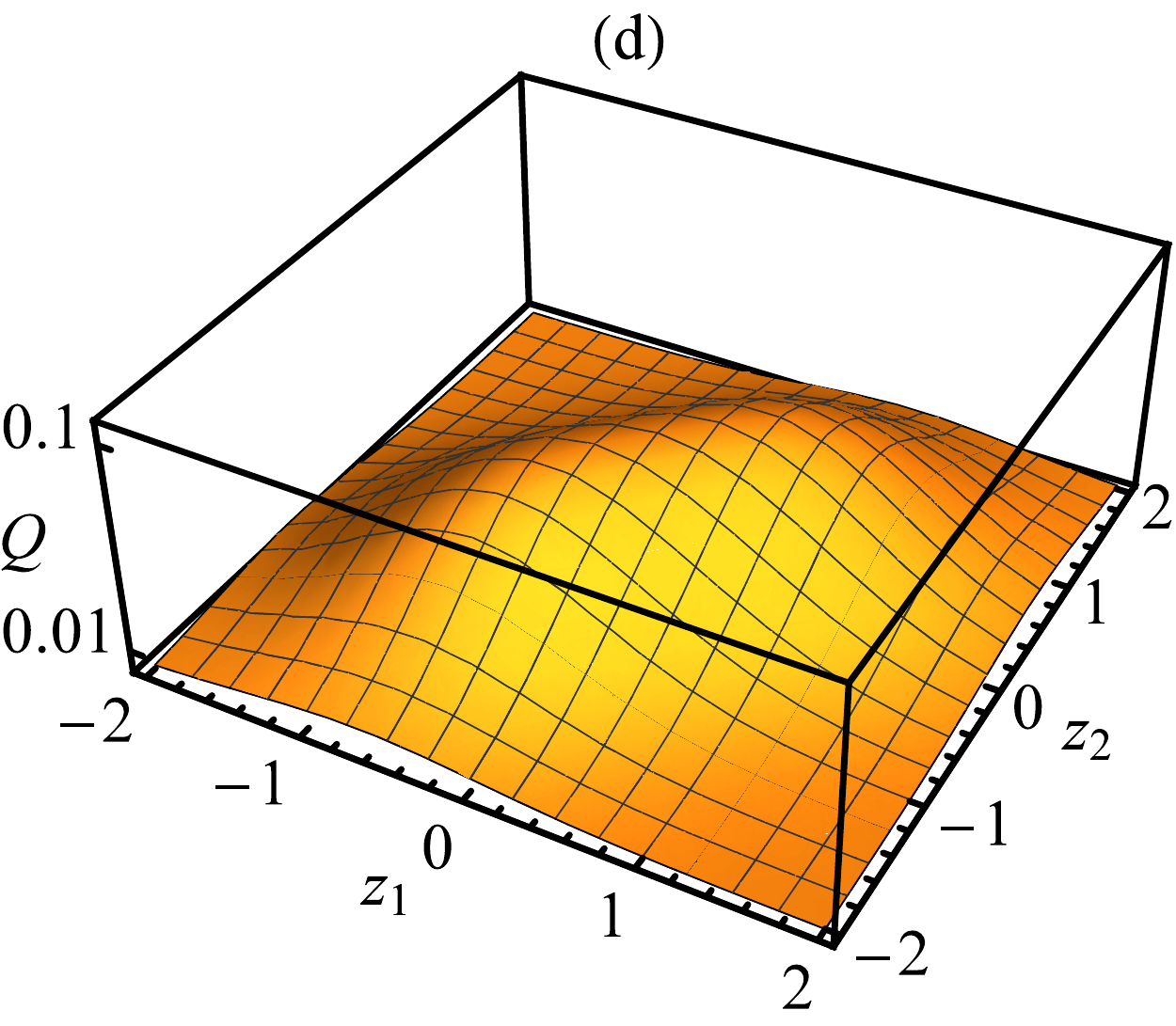}
\includegraphics[width=5.8cm]{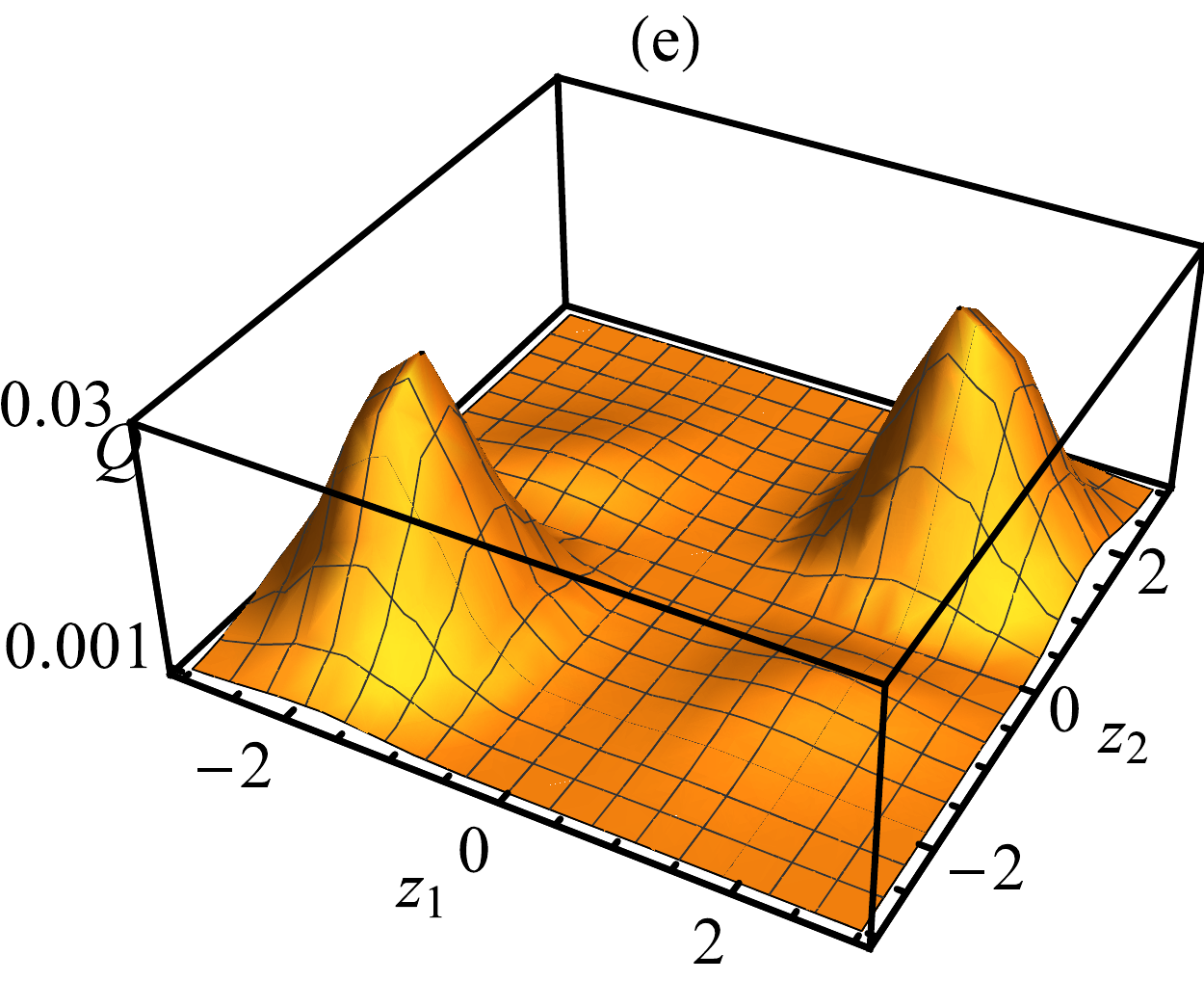}
\includegraphics[width=5.8cm]{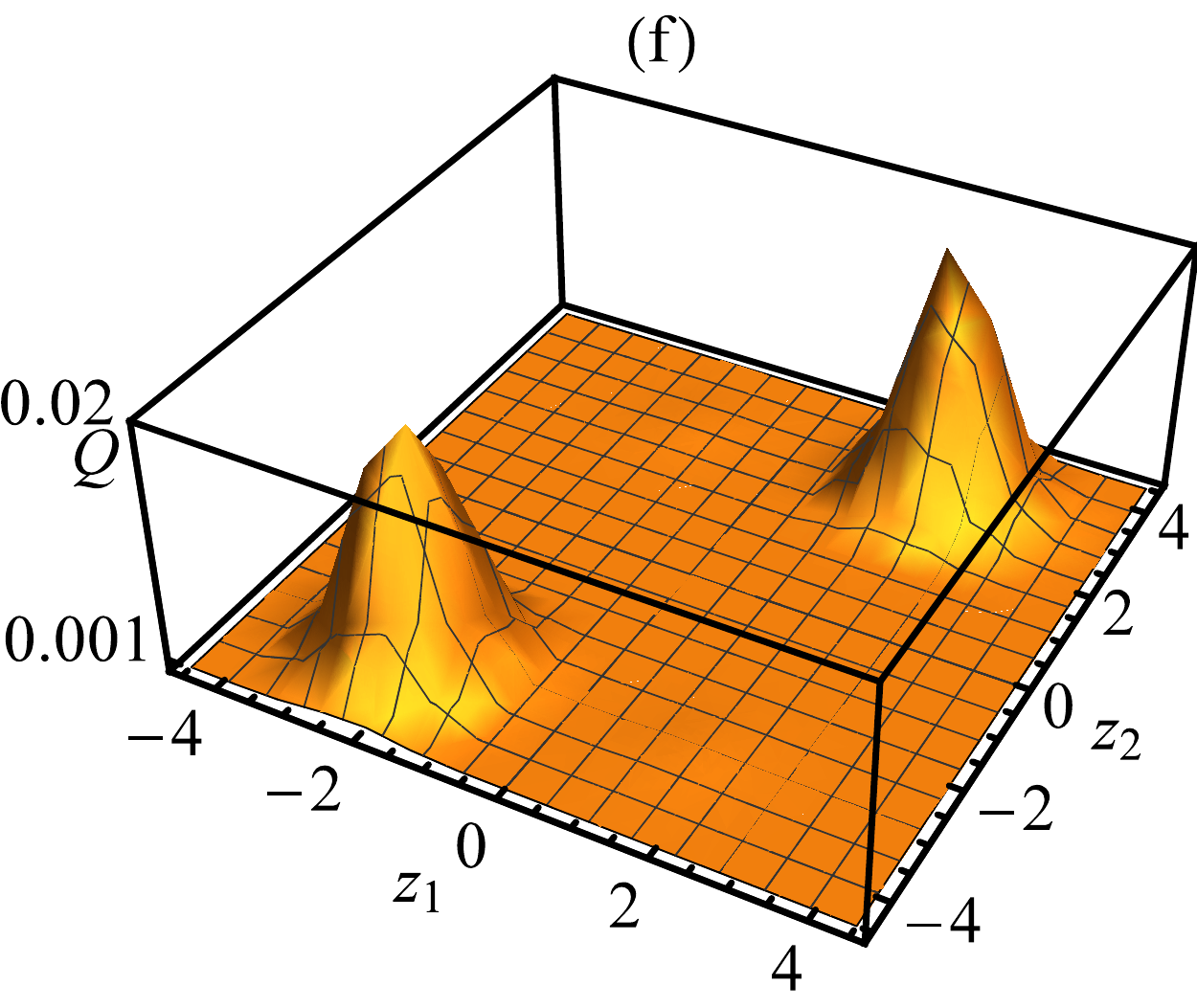}

\caption{(Color online) $Q$-function of $\ket{\psi_1^+(\alpha, m, n)}$ for different values of photon-excitations: $m=n=0$ in (a) and (d), $m=2$, $n=1$ in (b) and (e), $m=3$, $n=7$ in (c) and (f) respectively, and for fixed values of $|\alpha|^2$  ($|\alpha|^2=0.05$ in upper row and $|\alpha|^2=0.5$ in lower row).}
\label{fig3}
\end{figure*}

\begin{figure*}[htbp]
\centering
\includegraphics[width=5.7cm]{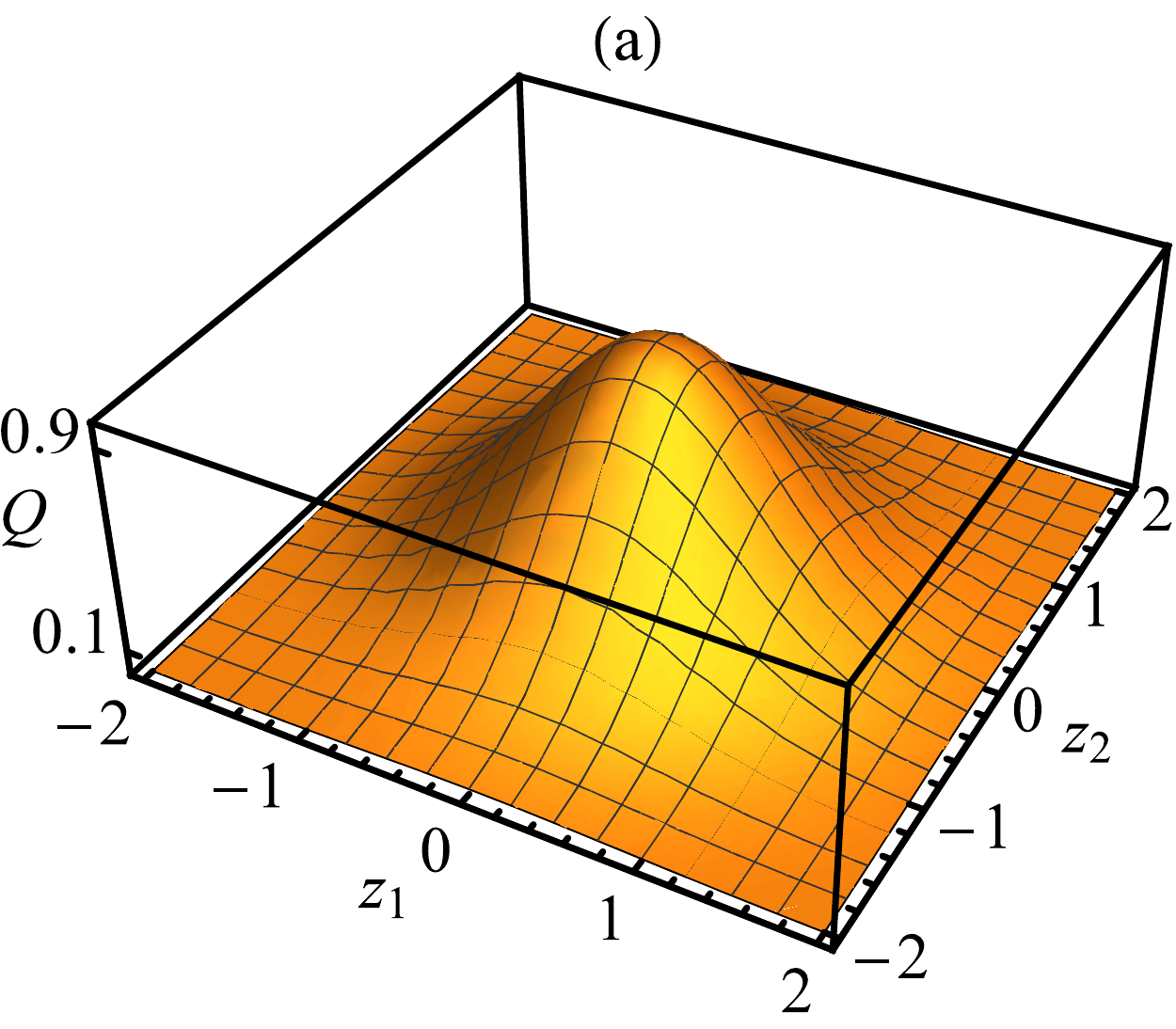}
\includegraphics[width=5.7cm]{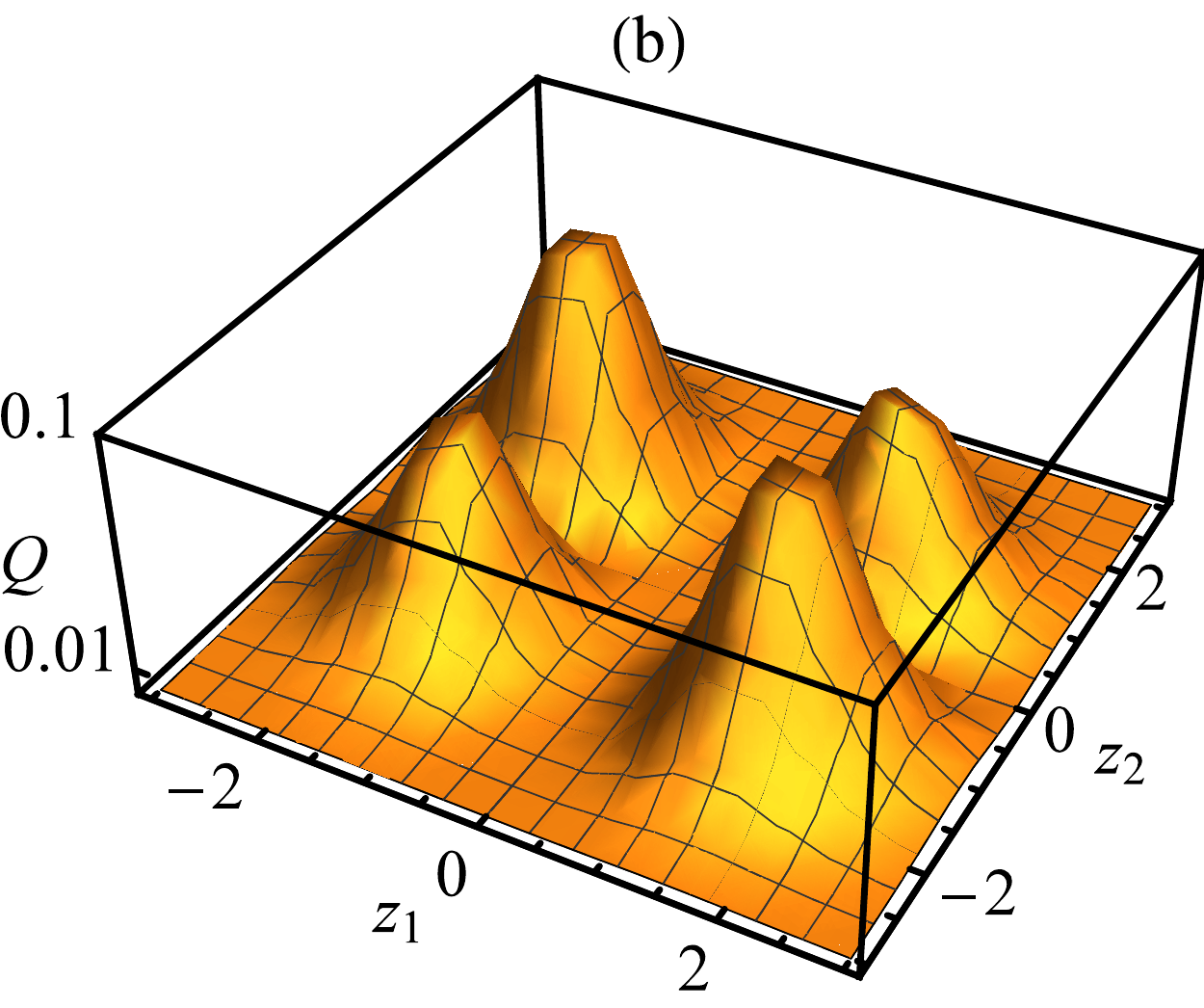}
\includegraphics[width=5.7cm]{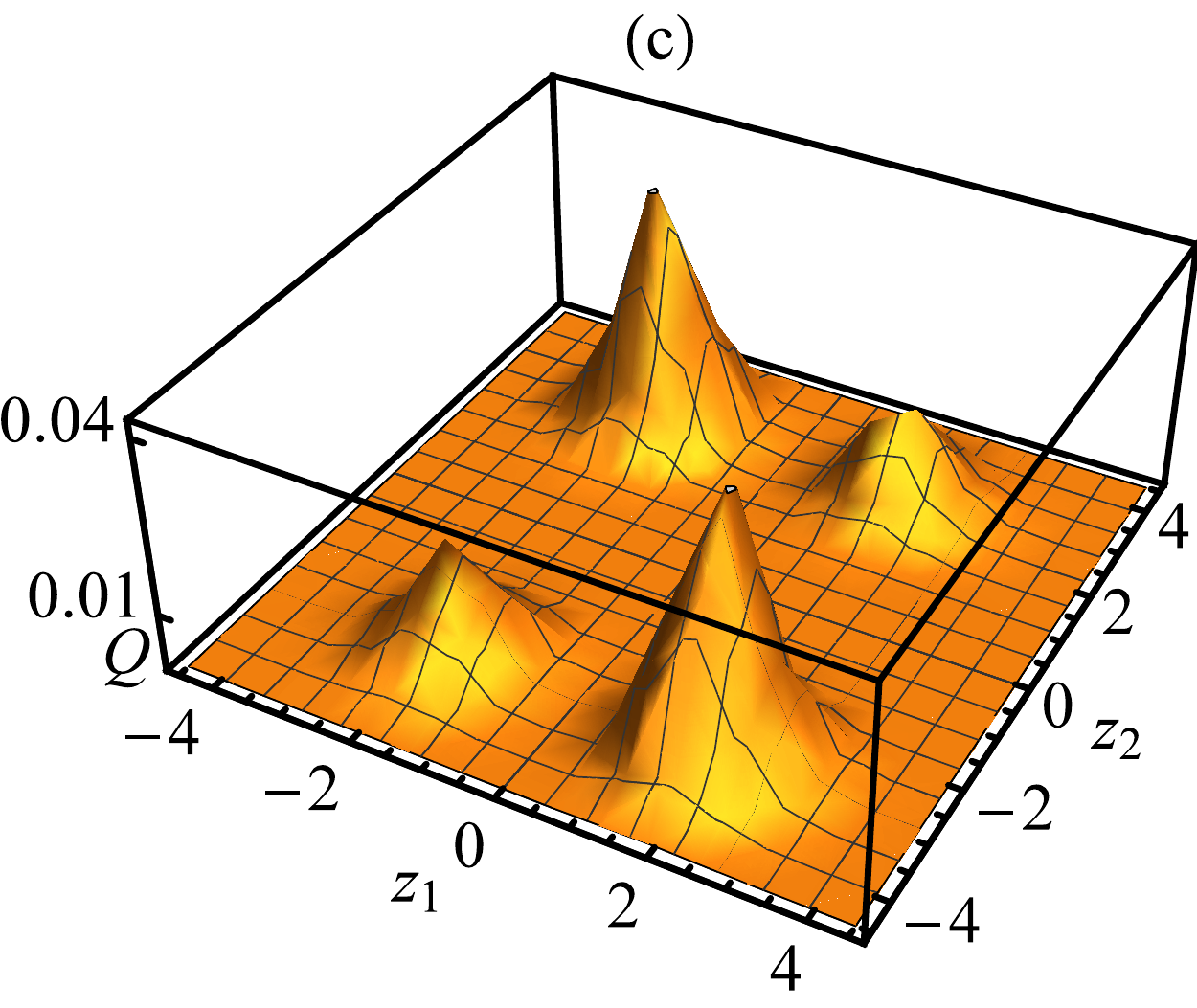}
\includegraphics[width=5.8cm]{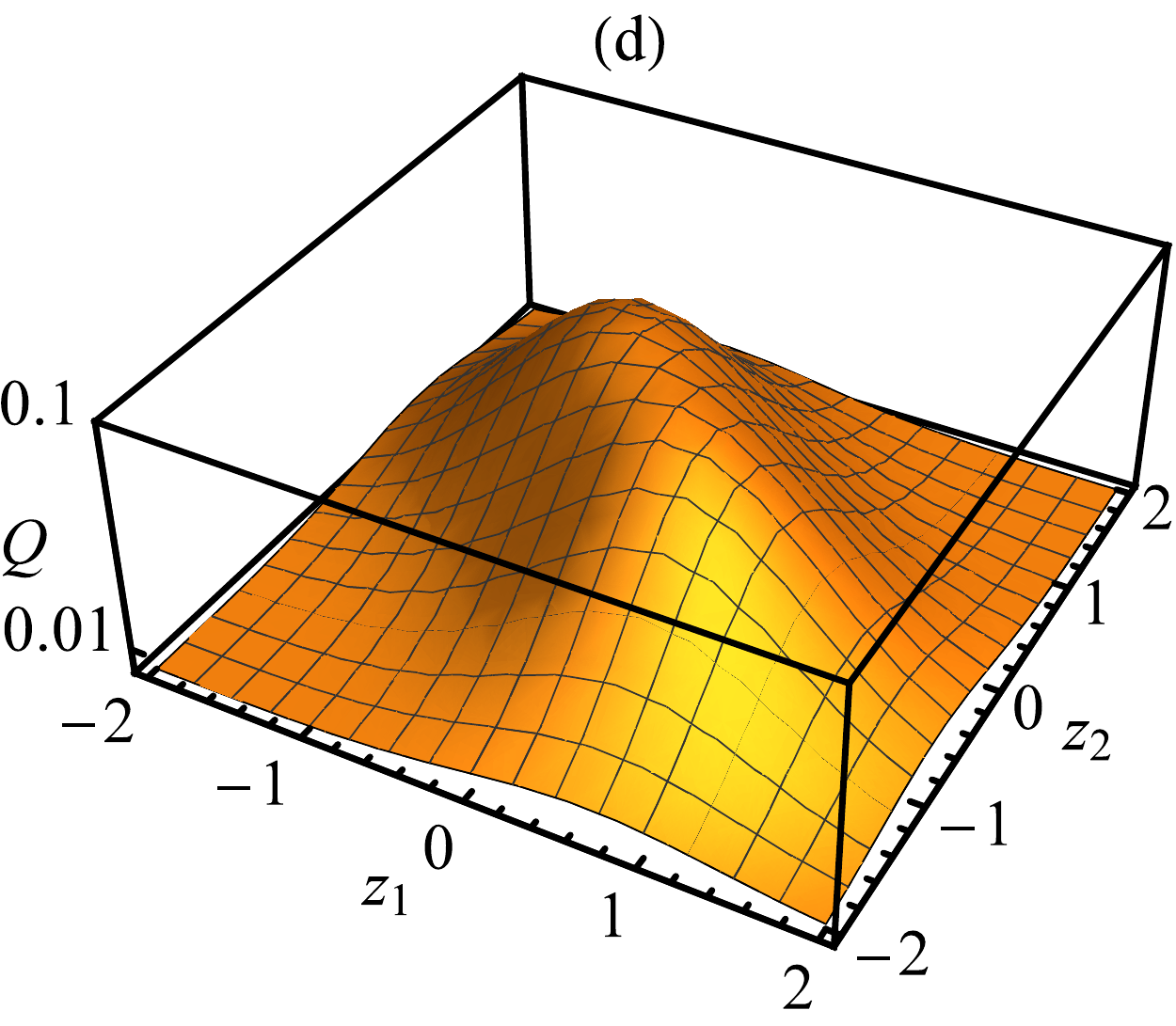}
\includegraphics[width=5.8cm]{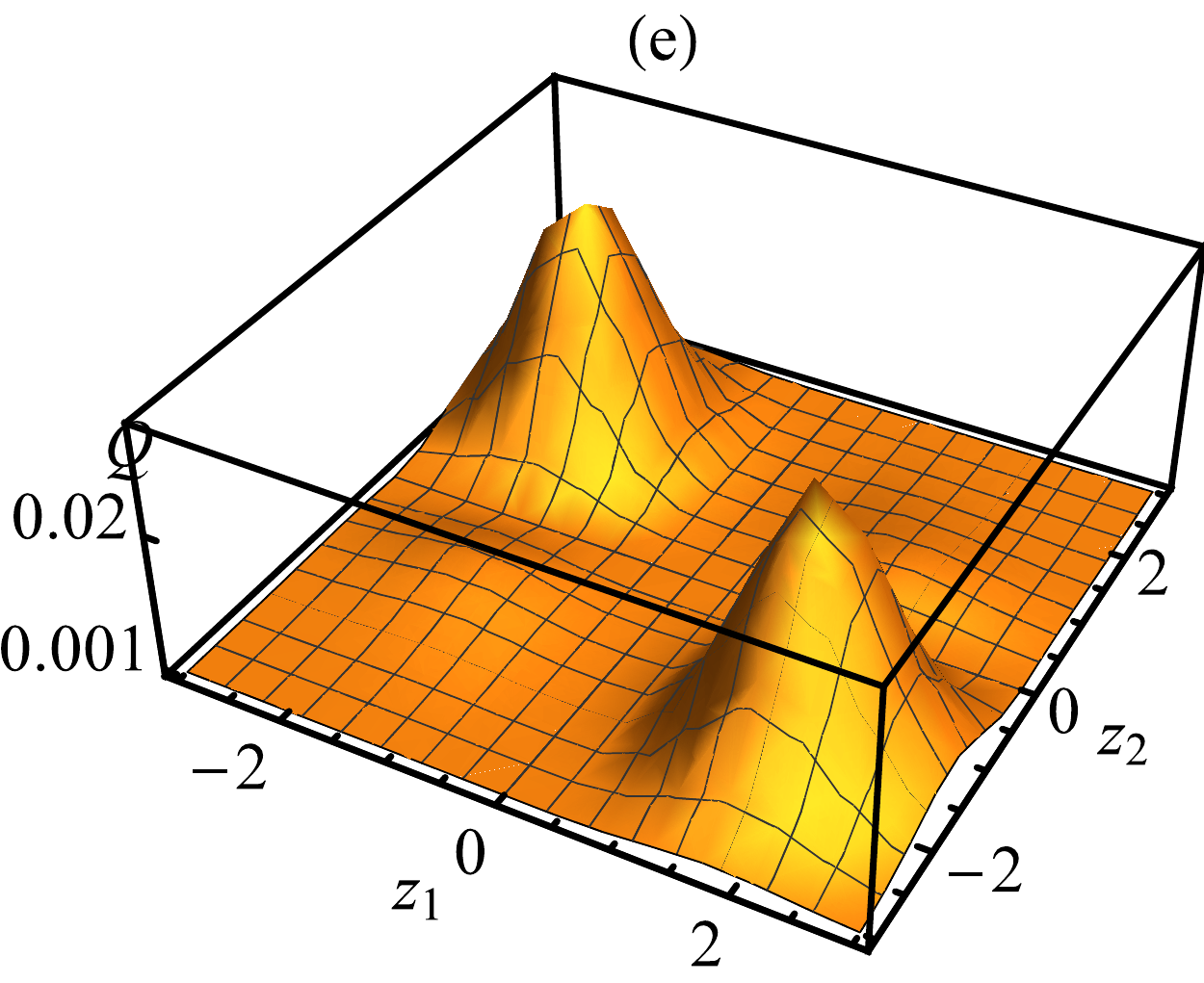}
\includegraphics[width=5.8cm]{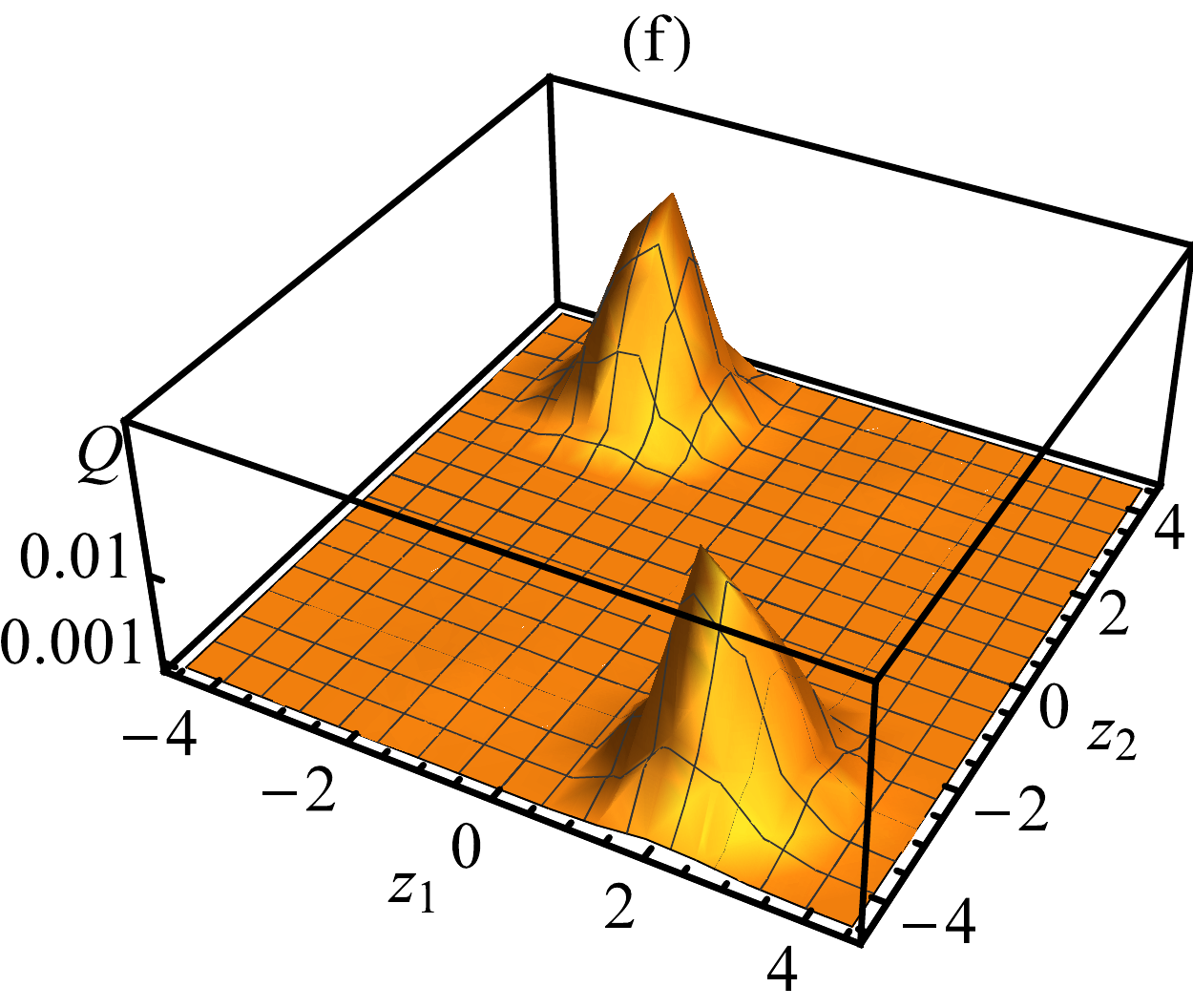}

\caption{(Color online) $Q$-function of $\ket{\psi_1^-(\alpha, m, n)}$ for different values of photon-excitations: $m=n=0$ in (a) and (d), $m=2$, $n=1$ in (b) and (e), $m=3$, $n=7$ in (c) and (f) respectively, and for fixed values of $|\alpha|^2$  ($|\alpha|^2=0.05$ in upper row and $|\alpha|^2=0.5$ in lower row).}
\label{fig4}
\end{figure*}

It is clearly seen that the $P$-function is highly singular, consisting of a series of terms of higher-order derivatives of a delta function.
There is another distribution function called the $Q$-function that is related to the $P$-function by \cite{mehta}
\begin{eqnarray}\nonumber
& & Q(z_1, z_2)\\\nonumber & = & \frac{1}{\pi^2}\bra{z_1, z_2}\rho\ket{z_1, z_2}\\
& = & \frac{1}{\pi^2} \int d^2\mu d^2\nu P(\mu, \nu)e^{-|\mu-z_1|^2-|\nu-z_2|^2}
\end{eqnarray}
which is always positive. Note that if the function $P(\mu,\nu)$ is like a classical probability distribution, then $Q(z_1, z_2) > 0$. However, if $Q$ is zero, then $P$ must become at least negative in some parts, which refers to the nonclassicality of the state. Hence, the exact zeros of the $Q$-function are also a signature for the nonclassicality of the field. The $Q$ functions of PAECSs can be calculated as
\begin{eqnarray}\nonumber
& & Q_{\psi_1^\pm}(z_1, z_2)\\\nonumber & = & \frac{1}{\pi^2}\bra{z_1, z_2}\rho_{\psi_1^\pm}\ket{z_1, z_2}\\
& = & \frac{(N_{mn}^\pm)^2}{\pi^2}|z_1|^{2m}|z_2|^{2n}e^{-2|\alpha|^2-|z_1|^2-|z_2|^2}\\\nonumber
& & \Big[e^{\xi^*\alpha+\xi\alpha^*}\pm e^{-\xi^*\alpha+\xi\alpha^*}\pm e^{\xi^*\alpha-\xi\alpha^*}+e^{-\xi^*\alpha-\xi\alpha^*} \Big]
\end{eqnarray}
and
\begin{eqnarray}\nonumber
& & Q_{\psi_2^\pm}(z_1, z_2)\\\nonumber & = & \frac{1}{\pi^2}\bra{z_1, z_2}\rho_{\psi_2^\pm}\ket{z_1, z_2}\\
& = & \frac{(N_{mn}^\pm)^2}{\pi^2}|z_1|^{2m}|z_2|^{2n}e^{-2|\alpha|^2-|z_1|^2-|z_2|^2}\\\nonumber
& & \Big[e^{\eta^*\alpha+\eta\alpha^*}\pm e^{-\eta^*\alpha+\eta\alpha^*}\pm e^{\eta^*\alpha-\eta\alpha^*}+e^{-\eta^*\alpha-\eta\alpha^*} \Big]
\end{eqnarray}
where $$\xi\equiv z_1+z_2,\,\,\eta\equiv z_1-z_2$$.

%\begin{figure}[htbp]
%\centering
%\includegraphics[scale=0.5]{qfig21.eps}
%\includegraphics[scale=0.5]{qfig22.eps}
%\includegraphics[scale=0.5]{qfig23.eps}
%
%\caption{(Color online) $Q$-function of $\ket{\psi_2^+(\alpha, m, n)}$ for a fixed value of $|\alpha|^2=0.04$ and for different values of photon-excitations: (a) $m=n=0$, (b) $m=1$, $n=0$, and (c) $m=0$, $n=1$, respectively.}
%\label{fig3}
%\end{figure}

To see the behavior of the $Q$-function of the PAECSs, we plot $Q(z_1, z_2)$ as a function of $z_1$ and $z_2$, for fixed values of $|\alpha|^2$. In the context of entropy of entanglement, $|\alpha|=0.2$ is a point of utmost interest as the behaviour of the entropy is just reversed before and after this point (see Figs.~\ref{fig1} and \ref{fig2}). To check if the $Q$ function is also responsive to this specific point, we assume $|\alpha|^2=0.05$ and $|\alpha|^2=0.5$ in the first and the second rows of the $Q$ plots, respectively. The choice of $|\alpha|$ is in accordance with the experimental values given by Zavatta et al. \cite{zavatta}. If there is no photon excitation, i.e., when $m = n = 0$ (see Figs.~\ref{fig2}(a) and ~\ref{fig3}(a)), $Q(z_1, z_2)$ of $\ket{\psi_1^+(\alpha, m, n)}$ exhibits a single peak structure at centre. If $m \neq 0$ and $n \neq 0$, the single peak is split into four adjacent peaks. While increasing the excitation photon numbers $m$ and $n$ further, these four peaks gradually move away from each other along the $z_1$ and $z_2$ directions, respectively. We can see that diagonal two peaks tend to be flatten with increasing values of $m$ and $n$. It is also observed that when $|\alpha|^2$ changes from 0.05 to 0.5, the peaks disappear immediately. In addition, to make a comparison of the entanglement properties between the PAECSs ($\ket{\psi_1^+(\alpha, m, n)}$ and $\ket{\psi_1^-(\alpha, m, n)}$), we plot $Q(z_1, z_2)$ of $\ket{\psi_1^-(\alpha, m, n)}$ in Fig.~\ref{fig4}. One can clearly see that the influence of photon excitations on $Q(z_1, z_2)$ of $\ket{\psi_1^-(\alpha, m, n)}$  is similar to that of $\ket{\psi_1^+(\alpha, m, n)}$, except that the increasing values of $(m, n)$ make the other two diagonal peaks smooth.

\section{Conclusion}

In summary, we consider a class of the photon-added entangled coherent states (PAECSs), which are
obtained by acting a creation operator on ECSs. We give the Schmidt decomposition of the PAECSs in
terms of the excited even (odd) coherent states. Then we calculate their von-Neumann entropy of entanglement and discuss the
influence of the photon excitations on quantum entanglement. We find that the entropy $E$ of PAECSs is sensitive to the
photon excitations to both modes. The line plots for entropy of $\ket{\psi_1^+(\alpha, m, n)}$ are increasing as $m$ changes from 0 to 20 and $n$ changes from 0 to 4, respectively. In case of $\ket{\psi_1^-(\alpha, m, n)}$ and $\ket{\psi_2^-(\alpha, m, n)}$, the states are always maximally entangled while $m = n$. We calculate the $Q$-distribution of the PAECSs and study its behavior graphically. The results show that the nonclassical effects revealed by this distribution function are affected by photon excitation numbers $m$ and $n$ in a significant way.

\end{document}